\documentclass[fleqn,usenatbib]{mnras}

\usepackage{newtxtext,newtxmath}

\usepackage{amsmath}
\usepackage{autobreak}
\usepackage{bm}
\usepackage[T1]{fontenc}

\DeclareRobustCommand{\VAN}[3]{#2}
\let\VANthebibliography\thebibliography
\def\thebibliography{\DeclareRobustCommand{\VAN}[3]{##3}\VANthebibliography}

\usepackage{graphicx}	
\usepackage{amsmath}	
\usepackage{amssymb}	
\usepackage[switch]{lineno}	

\defcitealias{2007SoPh..246..213A}{A07} 
\defcitealias{2011SSRv..158..289G}{GER11}
\defcitealias{2014ApJ...789...48O}{ORT14}
\defcitealias{1974Book..Whitham}{W74}
\defcitealias{2021SoPh..296...95Y}{Y21}

\renewcommand{\vec}[1]{\ensuremath{\bm{#1}}}

\newcommand{\uvec}[1]{\ensuremath{\bm{e}_{#1}}}
\newcommand{\Exp}[1]{\ensuremath{{\rm e}^{#1}}}

\newcommand{\Alf}{Alfv$\acute{\rm e}$n}
\newcommand{\Alfvenic}{Alfv$\acute{\rm e}$nic}

\newcommand{\va}{\ensuremath{v_{\rm A}}}
\newcommand{\vai}{\ensuremath{v_{\rm Ai}}}
\newcommand{\vae}{\ensuremath{v_{\rm Ae}}}

\newcommand{\rhoi}{\ensuremath{\rho_{\rm i}}}
\newcommand{\rhoe}{\ensuremath{\rho_{\rm e}}}

\newcommand{\vgy}{\ensuremath{v_{{\rm gr},y}}}
\newcommand{\vgz}{\ensuremath{v_{{\rm gr},z}}}

\newcommand{\vph}{\ensuremath{v_{\rm ph}}}
\newcommand{\vpy}{\ensuremath{v_{{\rm ph},y}}}
\newcommand{\vpz}{\ensuremath{v_{{\rm ph},z}}}

\title[3D Kink Waves in Solar Coronal Slabs]{Three-Dimensional Propagation of Kink Wave Trains in Solar Coronal Slabs}

\author[B. Li et al.]{
Bo Li,$^{1}$\thanks{E-mail: bbl@sdu.edu.cn}
Mingzhe Guo,$^{1}$
Hui Yu,$^{1}$
Shao-Xia Chen,$^{1}$
Mijie Shi $^{1}$
\\
$^{1}$Shandong Provincial Key Laboratory of Optical Astronomy and Solar-Terrestrial Environment,
   Institute of Space Sciences, Shandong University, Weihai 264209, China
}

\date{Accepted XXX. Received YYY; in original form ZZZ}

\pubyear{2022}

\begin{document}
\label{firstpage}
\pagerange{\pageref{firstpage}--\pageref{lastpage}}
\maketitle

\begin{abstract}
Impulsively excited wave trains are of considerable interest
    in solar coronal seismology.
To our knowledge, however, it remains to 
    examine the three-dimensional (3D) dispersive
    propagation of impulsive kink waves in straight, field-aligned, symmetric,
    low-beta, slab equilibria that are structured only in one transverse direction.
We offer a study here, starting with an analysis of linear oblique kink modes
    from an eigenvalue problem perspective. 
Two features are numerically found for continuous and step structuring alike,
    one being that the group and phase velocities may
    lie on opposite sides of the equilibrium magnetic field ($\vec{B}_0$), 
    and the other being that the group trajectories
    extend only to a limited angle from $\vec{B}_0$.
We justify these features by making analytical progress for the step structuring.
More importantly, we demonstrate by a 3D time-dependent simulation that 
    these features show up in the intricate interference patterns
    of kink wave trains that arise from a localized initial perturbation. 
In a plane perpendicular to the direction of inhomogeneity,
    the large-time slab-guided patterns 
    are confined to a narrow sector about $\vec{B}_0$,
    with some wavefronts propagating toward $\vec{B}_0$. 
We conclude that the phase and group diagrams 
    lay the necessary framework for understanding
    the complicated time-dependent behavior
    of impulsive waves. 
\end{abstract}




\section{Introduction}
\label{sec_intro}
The idea of solar coronal seismology 
   (SCS, e.g., \citealt{1984ApJ...279..857R,1970PASJ...22..341U})
   relies heavily on theoretical understandings
   of magnetohydrodynamic (MHD) waves in structured media. 
Consequently, there exists an extensive list of studies on 
   MHD waves in both slab and cylindrical equilibria from
   both eigenvalue problem (EVP) and initial value problem (IVP)
   standpoints
   \citep[see e.g.,][for reviews]{2000SoPh..193..139R,2005LRSP....2....3N,2020ARA&A..58..441N}.
However, wave trains impulsively excited by localized perturbations seem to be 
   under-examined \citep[see the reviews by e.g.,][]{2008IAUS..247....3R,2021SSRv..217...73N}, 
   despite their direct involvement in the establishment of SCS
   \citep{1983Natur.305..688R,1984ApJ...279..857R}. 
This is particularly true for impulsive kink waves, 
   for which a detailed IVP study
   was initiated for cylindrical equilibria only recently
   \citep[][hereafter \citetalias{2014ApJ...789...48O}]{2014ApJ...789...48O}.
Equally surprising is the apparent lack of a study on the three-dimensional (3D)
   propagation of impulsive kink waves in a slab equilibrium, despite the long-lasting
   interest in EVP studies on oblique kink modes in solar contexts
   \citep[e.g.,][]{1978ApJ...226..650I,1979ApJ...227..319W}
   and despite the literature on IVP studies in 2D
   \citep[e.g.,][]{1993SoPh..144..101M,2006SoPh..236..273O,2013A&A...560A..97P,2021MNRAS.505.3505K,2022MNRAS.515.4055G}.   
This manuscript aims at presenting such a study 
   with both an EVP (Section~\ref{sec_EVP}) 
   and an IVP (Section~\ref{sec_3Dnum}) approach.
We choose to leave out the observational implications,
   detailing instead how we       
   connect the EVP results, the group diagrams in particular,
   with our 3D simulation by the method of stationary phase
   \citep[MSP, Chapter~11 in][hereafter \citetalias{1974Book..Whitham}]{1974Book..Whitham}. 
The group diagrams of 3D kink modes are new to our knowledge,
   and so is the application of MSP to this context. 

\section{the EVP Perspective}
\label{sec_EVP}

This section works in zero-beta MHD, involved in which 
    are the mass density ($\rho$), velocity ($\vec{v}$), 
    and magnetic field $\vec{B}$. 
Let $(x, y, z)$ denote a Cartesian coordinate system, 
    and let the subscript $0$ denote the equilibrium quantities.
We consider only static equilibria ($\vec{v}_0 = 0$),
    and {take $\vec{B}_0$ to be} 
    $z$-directed and uniform ($\vec{B}_0 = B_0 \uvec{z}$). 
We assume that the equilibrium density ($\rho_0$) is an even function 
    of $x$, following 
\begin{equation}
\label{eq_rho_prof}
  \rho_0(x)
= \rhoe+\dfrac{\rhoi-\rhoe}{1+|x/d|^\alpha}. 
\end{equation} 
A continuous profile is chosen to comply with
    Section~\ref{sec_3Dnum}.
Here $d$ represents some 
    slab half-width, and $\alpha$ is some steepness parameter.
By ``internal'' and ``external'' we refer to 
    the equilibrium quantities at the slab axis ($x=0$, subscript~${\rm i}$) 
    and infinitely far ($|x|\to\infty$, subscript~${\rm e}$), respectively.
The internal (external) \Alf\ speed $\vai$ ($\vae$) 
    then derives from the internal density
    $\rhoi$ (external density $\rhoe$), following 
    $\va^2 = B_0^2/(\mu_0 \rho_0)$ with $\mu_0$ 
       the magnetic permeability of free space.
Only the half-plane $x\ge 0$ needs to be considered.
By ``out-of-plane'' we refer to the $y$-direction.        
By ``textbook'' we refer to the situation where ideal MHD is adopted,
    out-of-plane propagation is neglected, and
    $\rho_0(x)$ takes a step profile 
    \citep[$\alpha\to\infty$, see the textbook by][]{2019CUP_Roberts}. 
An infinity of branches of trapped kink modes arise in this case,
    and we label a branch by the transverse order $l=1,2,\cdots$
    \citep[e.g.,][Figure~2]{2018ApJ...855...53L}. 
{By ``kink modes'' we restrict ourselves
    to those that are physically connected
    to the $l=1$ textbook modes.} 
We additionally fix $[\rhoi/\rhoe, \alpha]$ at $[3, 10]$ 
    unless stated otherwise. 
           
\subsection{Eigenvalue Problem}
\label{sec_sub_EVP}

A linear analysis proves insightful.
For prescription~\eqref{eq_rho_prof}, however,
    oblique kink modes are in general resonantly absorbed
    in the \Alf\ continuum unless $\alpha \to \infty$
    (see the review by \citealt[][hereafter \citetalias{2011SSRv..158..289G}]{2011SSRv..158..289G}; also the earliest studies
    by e.g., \citealt{1973ZPhy..261..203T,1974PhRvL..32..454H}).
We adopt a resistive eigenmode approach
    (see \citetalias{2011SSRv..158..289G} for conceptual clarifications).
Let the subscript~$1$ denote small-amplitude perturbations, which
    are Fourier-decomposed as
\begin{equation}
\label{eq_Fourier}
   f_1 (x, y, z; t)
 = \Re\{\tilde{f}(x)\exp[-\imath (\Omega t- k_y y -k_z z)]\},
\end{equation}
   where $\Omega$ is the angular frequency, and $k_z$ ($k_y$)
   the real-valued axial (out-of-plane) wavenumber.
A much-studied EVP ensues in linear resistive MHD,
   involving only $\tilde{v}_x$, $\tilde{v}_y$, 
   $\tilde{B}_x$, $\tilde{B}_y$, and $\tilde{B}_z$
   (\citealt{1992SoPh..138..233G}; \citealt{1995JPlPh..54..129R};
    \citealt{2007SoPh..246..213A}, \citetalias{2007SoPh..246..213A}).
The governing equations   
   are {identical to Equations~(6) to (10) in 
   \citet[][\citetalias{2021SoPh..296...95Y}]{2021SoPh..296...95Y}}.
As in \citetalias{2021SoPh..296...95Y},   
   kink eigensolutions are guaranteed by the boundary condition 
   at $x=0$, namely
    $d\tilde{v}_x/dx= \tilde{v}_y=d\tilde{B}_x/dx=\tilde{B}_y=\tilde{B}_z=0$. 
We require that all Fourier amplitudes vanish at infinity. 

The kink eigensolution of interest is unique in that
   its damping rate becomes independent 
   of the electric resistivity $\eta$ when $\eta$ is small enough
   \citep[see][for the first demonstration]{1991PhRvL..66.2871P}.
Its $\eta$-independent eigenfrequency {is formally expressible as} 
\begin{equation}
\label{eq_omega_formal}
  \dfrac{\Omega d}{v_{\rm Ai}} 
= \mathcal{W}
  \left(
       k_y d, k_z d~\left|~
       \dfrac{\rhoi}{\rhoe}, \alpha\right.
  \right)
= \mathcal{W}(k_y d, k_z d).
\end{equation}
The second equal sign emphasizes our focus on 
   the $k_y$- and $k_z$-dependencies.
We follow \citetalias{2021SoPh..296...95Y} to 
   numerically establish $\mathcal{W}$ by {solving the EVP} with the PDE2D code
   (\citealt{1988Sewell_PDE2D}; see \citealt{2006ApJ...642..533T}
   for its first solar application).       
Let $\omega$ ($\gamma$) denote the real (imaginary) part of $\Omega$.
Only damping eigensolutions are sought ($\gamma < 0$).
Let asterisks denote complex conjugate.      
The governing equations then dictate
   that if $\Omega$ is an eigenfrequency, then 
   so is $-\Omega^{*}$.
Likewise, if $\Omega$ is an eigenfrequency for a given $[k_y, k_z]$,
   then it remains {so} for $[-k_y,k_z]$, $[k_y,-k_z]$,
   and $[-k_y,-k_z]$.
It therefore suffices to assume $\omega>0$ and consider only 
   the quadrant $k_y \ge 0$, $k_z>0$.

\begin{figure}
\centering
\includegraphics[width=.99\columnwidth]{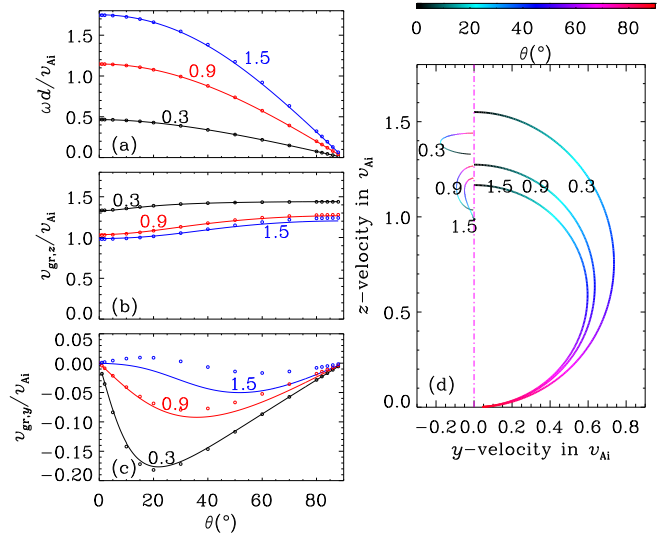}
\caption{
Dispersion properties of oblique kink modes in a 
    slab equilibrium structured only in the $x$-direction,
    the density contrast (steepness parameter) being
    $\rhoi/\rhoe=3$ ($\alpha=10$).
The obliqueness is measured by the angle $\theta$ between 
    the equilibrium magnetic field $\vec{B}_0 = B_0 \uvec{z}$
    and a 2D wavevector $\vec{k}=k_y \uvec{y} + k_z \uvec{z}$. 
Left column: Dependencies on $\theta$ of 
        (a) the oscillation frequency $\omega$,
        (b) the $z$-component of the group velocity $\vgz$,
    and (c) the $y$-component $\vgy$.
Several values of $kd$ are examined as labeled. 
The results for a step density profile ($\alpha \to \infty$)
    are represented by the symbols for comparison. 
Right column: Phase (the thick curves) and group (thin)
    diagrams for a number of $kd$ as labeled. 
Both diagrams are color-coded by $\theta$. 
The dash-dotted line represents the vertical axis in
    the velocity plane. 
See text for more details.                
 }
\label{fig_vRainbow} 
\end{figure}

\subsection{Phase and Group Diagrams}
\label{sec_sub_groupV}
This subsection presents the phase and group diagrams.
We start by defining a 2D wavevector 
   $\vec{k}\coloneqq k_y\uvec{y}+k_z\uvec{z}$, where $k_y\ge 0$
   and $k_z>0$.
The wavevector is alternatively represented by $k$ and $\theta$, with 
   $k=|\vec{k}|$ and $\theta=\angle(\vec{k},\vec{B}_0)$.
We define the phase velocity 
   as $\vec{v}_{\rm ph}=(\omega/k)\uvec{k}$ 
   with $\uvec{k}$ being
   the unit vector along $\vec{k}$. 
The group velocity, on the other hand,
   is defined as 
   $\vec{v}_{\rm gr} = \vgy \uvec{y} + \vgz \uvec{z}$ with
         $\vgy = \partial\omega/\partial k_y$ 
   and   $\vgz = \partial\omega/\partial k_z$.        
Equation~\eqref{eq_omega_formal} then enables one to 
   convert the $k_y$- and $k_z$-dependencies of $\omega=\Re\Omega$
   into the $k$- and $\theta$-dependencies of $\vec{v}_{\rm gr}$, yielding
\begin{equation}
\label{eq_formal_vgyvgz}
\vgy/\vai = \mathcal{U}_y(kd, \theta), \quad 
\vgz/\vai = \mathcal{U}_z(kd, \theta).
\end{equation}

We largely focus on how $\vec{v}_{\rm ph}$ or $\vec{v}_{\rm gr}$
   varies with $\theta$, for which purpose 
   $\omega$, $\vgz$ and $\vgy$ for the chosen
   $[\rhoi/\rhoe, \alpha] = [3, 10]$
   are plotted against $\theta$
   by the solid curves 
   in Figures~\ref{fig_vRainbow}a to \ref{fig_vRainbow}c. 
A number of $kd$ are examined as labeled. 
The step results ($\alpha\to\infty$)
      are additionally presented by the symbols. 
One sees that the finite $\alpha$ curves differ
    appreciably from the symbols only 
    for $\vgy$ when $kd = 1.5$.
This is understandable because kink modes with larger $k$ possess
    shorter spatial scales, and therefore sense
    the finite inhomogeneity scale more readily.
Our point, however, is that the finite $\alpha$ results
    can be largely understood with the step ones, for which
    some analytical progress is possible. 
This practice proves necessary given the rather involved 
    $\theta$-dependencies, those of $\vgy$ in particular. 
We proceed to define
\begin{equation}
\label{eq_def_kappaie}
  \kappa^2_{\rm i, e} 
= k_z^2         -\dfrac{\omega^2}{v^2_{\rm Ai, e}},  
  \quad 
  m^2_{\rm i, e} 
= k_y^2 + \kappa^2_{\rm i, e}, 
\end{equation}
   where $m_{\rm i, e}$ acts as some effective $x$-wavenumber
   for oblique kink modes (e.g., Equation~(15) in \citetalias{2021SoPh..296...95Y}).
We note that $m_{\rm e}^2$ is positive by construction, 
   whereas $m_{\rm i}^2$ may be negative for small $k_y$
   (\citetalias{2007SoPh..246..213A}, Figure~2).
We take $\arg m_{\rm i}=\pi/2$ when $m_{\rm i}^2<0$ without loss of generality.   
A dispersion relation (DR) then writes
   (e.g., \citetalias{2007SoPh..246..213A,2021SoPh..296...95Y})    
\begin{equation}
\label{eq_DR_step}
   \coth(m_{\rm i} d) 
= -\dfrac{\kappa_{\rm i}^2}{\kappa_{\rm e}^2} 
   \frac{m_{\rm e}}{m_{\rm i}}.   			
\end{equation}

The dispersion behavior at large $\theta$ can be explained
    with Equation~\eqref{eq_DR_step} by assuming 
    $m_{\rm i}\approx m_{\rm e}\approx k_y$. 
It was first shown by \citet{1998JPSJ...67.2322T}
    in fusion contexts that
\begin{equation}
\label{eq_DR_step_largeKY}
        \omega 
\approx k_z 
        C(k_y), 
\quad         
        C^2(k_y)
=       \vai^2 
        \dfrac{1+\tanh(k_y d)}{\rhoe/\rhoi+\tanh(k_y d)},
\end{equation}
    which was put in current form by \citetalias{2021SoPh..296...95Y}.
Note that Equation~\eqref{eq_DR_step_largeKY} can also be found by imposing
    the incompressibility condition from the outset, with the physical reasons well
    documented for cylindrical equilibria
    \citep[e.g.,][]{2009A&A...503..213G,2012ApJ...753..111G,2015ApJ...803...43S}. 
Note further that Equation~\eqref{eq_DR_step_largeKY} leads to   
    $\kappa^2_{\rm e} < |\kappa^2_{\rm i}|$, 
    meaning that its range of validity $k_y^2 \gg |\kappa^2_{\rm i, e}|$
    becomes
\begin{equation}
\label{eq_DR_step_largeKY_rngVld}
k_y^2 [\rhoe/\rhoi+\tanh(k_y d)] \gg k_z^2 (1-\rhoe/\rhoi). 
\end{equation} 
Evidently, $\omega \to 0$ and hence $\vph \to 0$ when $\theta \to 90^\circ$.
Furthermore, one recognizes that $C(k_y)$ 
   decreases monotonically with $k_y$, meaning that
   $c_{\rm k}=\vai\sqrt{2/(1+\rhoe/\rhoi)}<C(k_y)<\vai\sqrt{\rhoi/\rhoe}=\vae$ with $c_{\rm k}$
      being the kink speed.
Consequently, the asymptotic value of $\vgz$ at $\theta \to 90^\circ$
   decreases monotonically with $k$, lying between $c_{\rm k}$ and $\vae$.
Likewise, $\vgy$ approaches zero from below.  
These step expectations are reproduced exactly by
   the symbols, explaining the finite $\alpha$ results
   almost quantitatively as well. 

We further employ Equation~\eqref{eq_DR_step} to 
    understand how $\vgy$ behaves at small $\theta$. 
Let $\hat{\omega}$ denote $\mathcal{W}(k_y=0, k_z)$
    (see Equation~\eqref{eq_omega_formal} with $\alpha\to\infty$). 
Consider a slightly different pair $[k_y, k_z]$ with
   $k_y^2/k^2_z \ll 1$.
We see only $k_y$ as variable, and Taylor-expand all terms
   in Equation~\eqref{eq_DR_step} about $k_y=0$.
Some lengthy algebra yields that
\begin{equation}
\label{eq_DR_step_smallKY}
        \omega(k_y, k_z)
\approx \hat{\omega} 
		\left[1-\dfrac{1}{2}
		        \dfrac{1/(\hat{\kappa}_{\rm e} d)-1}
		               {(\hat{\omega}d/\vai)^2+(k_z d)^2/(\hat{\kappa}_{\rm e} d)}  
		        (k_y d)^2
		\right],      
\end{equation}
   with $\hat{\kappa}_{\rm e}$ defined by
   $\hat{\kappa}^2_{\rm e} = k_z^2 - \hat{\omega}^2/\vae^2$.
Note that $\hat{\kappa}^2_{\rm e}>0$.   
Equation~\eqref{eq_DR_step_smallKY} 
   indicates that the $k_y$-correction is only quadratic, meaning
   that $\vgy\to 0$ when $\theta\to 0$.
For a fixed small~$\theta$, one further deduces that
   $\vgy$ remains negative for small $kd$ until reversing its sign
   when $kd$ exceeds some critical value.
To explain this, we recall the textbook result that
   $\hat{\omega}/k_z$  
   decrease monotonically with $k_z$
   from $\vae$ at $k_z d\to 0$ toward 
   $\vai$ when $k_z d\to\infty$ \citep[e.g.,][Section 5.5.5]{2019CUP_Roberts}. 
Now that
   $(\hat{\kappa}_{\rm e}d)^2 =(k_z d)^2 [1-\hat{\omega}^2/(k_z\vae)^2)]$,
   one recognizes that $\hat{\kappa}_e d \ll 1$ when $k_z d \ll 1$
   but increases monotonically with $k_z d$ toward large values 
   when $k_z d \gg 1$.
This necessarily changes the sign of $1/(\hat{\kappa}_{\rm e}d)-1$ at some $k_z d$,
   and hence a change of sign of $\vgy$. 
Both the curves and symbols
   in Figure~\ref{fig_vRainbow}c for small $\theta$ agree 
   with the analytical expectations. 
Somehow different between the step and finite $\alpha$ results is that    
   $kd=1.5$ is large enough to reverse the sign of $\vgy$ 
   for the former but not for the latter. 

Figure~\ref{fig_vRainbow}d gathers the finite $\alpha$ results
   to produce the phase and group diagrams,
   namely the trajectories that $\vec{v}_{\rm ph} = (\vpy,\vpz)$ (the thick curves)
   and $\vec{v}_{\rm gr}=(\vgy,\vgz)$ (thin) traverse
   when $\theta$ varies. 
These curves are color-coded by $\theta$ and labeled by $kd$. 
A primary result of this study, 
   Figure~\ref{fig_vRainbow}d is striking in that 
   the trajectories for any examined $kd$ are morphologically similar 
   to slow waves in a uniform low-beta MHD medium, 
   despite the absence of slow waves in zero-beta MHD
   (see, e.g., Figure~5.4 in \citealt{2019CUP_goedbloed_keppens_poedts}).
By ``similar'' we specifically emphasize that $\vec{v}_{\rm ph}$ and $\vec{v}_{\rm gr}$
   lie astride $\vec{B}_0 = B_0 \uvec{z}$, 
   and the group trajectories extend only to a limited angle from $\vec{B}_0$.
We deem it important to explore the physical reasons that yield the 
   peculiar group trajectories from the perspective of, say,
   restoring forces.
Equally important is to explore how the group trajectories behave in other
   configurations, one example being those that are associated with a magnetic shear
   ($\vec{B}_0 = B_{0y}(x)\uvec{y}+B_{0z}(x)\uvec{z}$;
    see e.g., \citealt{1974PhFl...17.1399C,2003A&A...402.1129A} 
    for some motivating ideas).
These explorations are nonetheless left for a future work.

\begin{figure}
\centering
\includegraphics[width=.98\columnwidth]{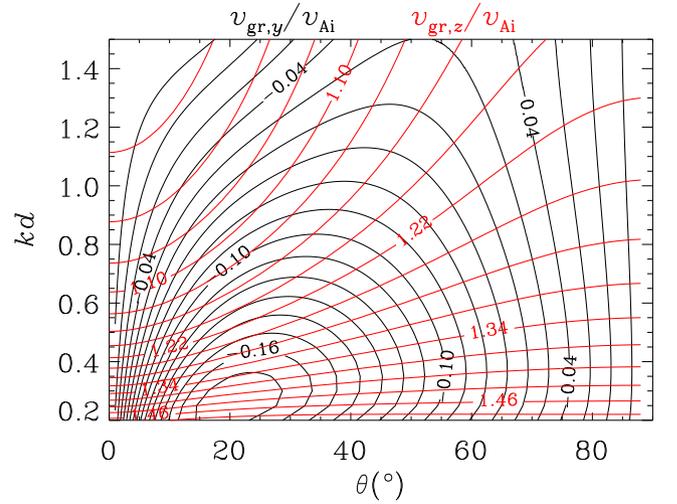}
\caption{
Distributions in the $k-\theta$ plane of 
   the $y$- and $z$-components of the group velocity,
   namely $\vgy$ (the black contours) and $\vgz$ (red), 
   for kink modes in a slab equilibrium with 
   $[\rhoi/\rhoe, \alpha] = [3, 10]$. 
 }
\label{fig_vgContour} 
\end{figure}

Figure~\ref{fig_vgContour} further surveys an extensive set of $[kd, \theta]$,
   presenting $\vgy$ and $\vgz$ as equally spaced contours colored black
   and red, respectively. 
One sees that $\vgy$ for a given $kd$ consistently approaches zero when 
   $\theta\to 0$ or $\theta\to 90^\circ$, thereby attaining a
   local minimum at some $kd$-dependent angle $\theta_{\rm min}$.
This $\theta_{\rm min}$ increases monotonically with $kd$,
   and $\vgy$ is subject to a global minimum in the examined range of $kd$
   (see the lower-left portion).      
Regarding $\vgz$, one sees that the mononotonic $\theta$-dependence 
   for a given $kd$ in Figure~\ref{fig_vRainbow}b actually persists. 
However, the upper-left corner indicates that
   $\vgz$ for a small $\theta$ will possess a nonmonotonic
   $kd$-dependence when $kd$ further increases.
This is indeed true.    
We avoid this complication, for 
   it is not specific to oblique propagation but
   well known for $\theta=0$ when $\alpha\to\infty$
   \citep[e.g.,][]{1995SoPh..159..399N}. 
Rather, by Figure~\ref{fig_vgContour} we stress that 
   inverting Equation~\eqref{eq_formal_vgyvgz} with a given pair
   $[\vgy/\vai,\vgz/\vai]$ does not yield a unique pair $[kd, \theta]$. 

\section{Time-Dependent Simulation}
\label{sec_3Dnum}
This section examines 3D impulsively excited kink waves by
    numerically evolving the ideal MHD equations with
    the MPI-AMRVAC code \citep{2018ApJS..234...30X}.
We prescribe the equilibrium density $\rho_0$ by
    Equation~\eqref{eq_rho_prof} with $[\rhoi/\rhoe,\alpha]=[3,10]$. 
A uniform temperature $T_0$ is specified 
    for simplicity, yielding a plasma beta of $0.021$ at $x=0$.
We choose a $z$-directed magnetic field $\vec{B}_0$ whose 
    magnitude varies with $x$ to
    maintain transverse force balance. 
This $x$-variation is nonetheless very weak, with $B_0$ at large $x$ 
    being larger than that at $x=0$ by only $\sim 0.7\%$.
Kink waves are excited by a perturbation
\begin{eqnarray}
  v_x(x,y,z; t=0)
= v_{\rm ini}\exp\left(-\dfrac{x^2}{2\sigma_x^2}\right)
             \exp\left(-\dfrac{y^2}{2\sigma_y^2}\right)
             \exp\left(-\dfrac{z^2}{2\sigma_z^2}\right),
\label{eq_vini}
\end{eqnarray}
    for which the spatial extent is chosen to be
    $\sigma_x=\sigma_y=\sigma_z=\sqrt{2}d$
    and the magnitude $v_{\rm ini}$ is set to be $0.1\vai$.  
Note that the same equilibrium was employed in the 2D study
    by \citet{2022MNRAS.515.4055G}, and Figure~4 therein demonstrated that
    a magnitude $v_{\rm ini}=0.1\vai$ ensures a linear behavior
    for the resulting kink wave train.
No nonlinearity is discerned in this 3D study either, which is understandable
    given that nonlinearity is weaker with the introduction of the third dimension.   

Our numerical setup is as follows. 
A subdomain $[-50,50]d\times[0,200]d\times[0,200]d$ of the full space
    is employed from symmetry considerations,
    with symmetric boundary conditions (BCs) specified
    at $y=0$ and $z=0$.  
Outflow BCs are implemented for the rest of the boundaries,
    where no spurious wave reflection is discerned.  
We adopt the second-order HLLD solver and the Woodward slope limiter
    when evaluating inter-cell fluxes,
    and choose the midpoint method for time marching
    with a Courant number of $0.5$.
A base grid of $64\times128\times128$ is adopted in $(x,y,z)$.
Four levels of adaptive mesh refinement (AMR)
    are implemented as triggered by density and velocity variations,
    resolving scales down to $\sim 0.19d$.
Despite the symmetry considerations and the implementation of AMR,     
    this simulation remains computationally expensive,
    and the chosen $\alpha=10$
    is the largest we can afford to avoid spurious waves emanating
    from the slab boundaries.

\begin{figure}
\centering
\includegraphics[width=.95\columnwidth]{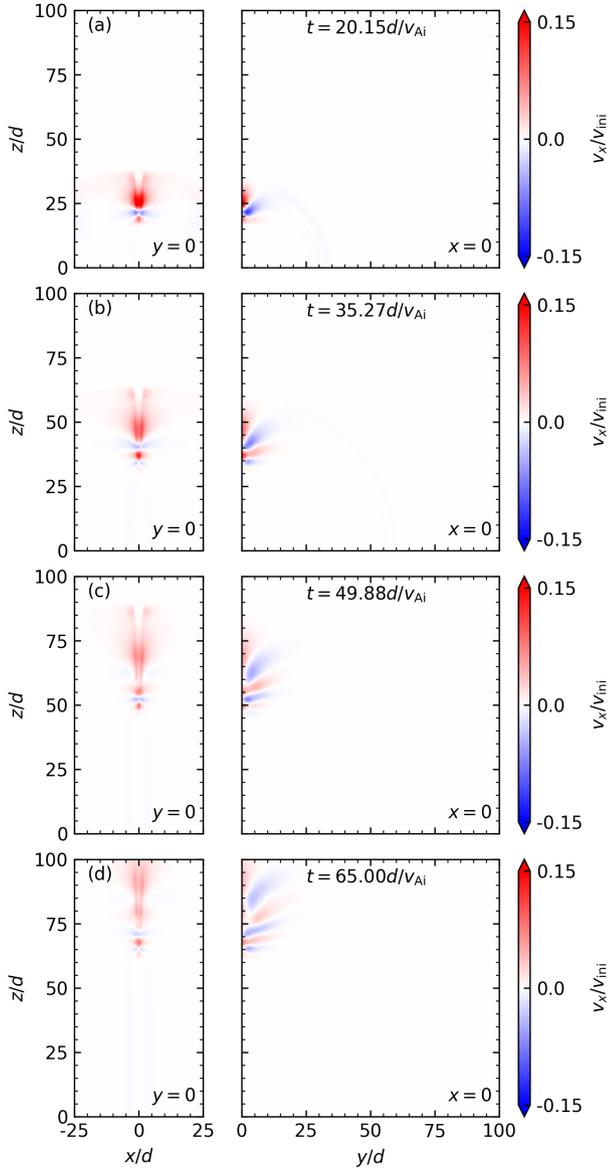}
\caption{
Snapshots of $v_x$ 
   in the $y=0$ (the left column) and $x=0$ (right) planes. 
These images are extracted from the animation
   attached to this figure.    
 }
\label{fig_vxAnim} 
\end{figure}

Figure~\ref{fig_vxAnim} presents several snapshots
    of $v_x$ in the $y=0$ (the left column)
    and $x=0$ planes (right). 
The $y=0$ cuts, together with the associated animation,
    make it clear that the signals comprise a slab-guided component
    and a laterally propagating component. 
This lateral component, manifested as a series of circular ripples,
    attenuates rather rapidly and becomes difficult to discern
    when $t\gtrsim 30d/\vai$.            
In contrast, the guided component persists throughout,
    characterized by that more ripples emerge with time
    and that small-scale variations tend to lag behind larger-scale ones. 
Two features arise in the $x=0$ cuts.
First, the nominal lateral component is present as well, despite
    that the $x=0$ plane is inside the slab.
Second, the guided component is confined to a rather narrow sector 
    about $\vec{B}_0$, encompassing multiple stripes that form an intricate 
    interference pattern. 

We focus on the $x=0$ cuts, for the $y=0$ patterns are
    largely 2D \citep[][Figure~4]{2022MNRAS.515.4055G}.
Neglecting finite pressure, 
   we assume $\alpha\to\infty$ to make quantitative progress.
One may write
\begin{eqnarray}
&&   v_x(x,y,z,t) = 
  \int_{-\infty}^{\infty} dk_y
  \int_{-\infty}^{\infty} dk_z  \nonumber \\
&&  \left\{
     \sum\limits_{j}\left[\mathcal{F}_j(x;k_y, k_z)
      \Exp{\imath\left(\omega_j t-k_y y-k_z z\right)}
      \right]
     +{\rm improper}
 \right\}, \label{eq_vx_IVPformal}     
\end{eqnarray}
   by drawing analogy with the cylindrical study by
   \citetalias{2014ApJ...789...48O} (see also \citealt{2022ApJ...928...33L}).
A spectral solution to the IVP, Equation~\eqref{eq_vx_IVPformal}
   means that all values of $k_y$ and $k_z$ are involved given
   the localization of the initial perturbation. 
The summation collects all trapped modes, 
   for which the frequency $\omega_j=\omega_j(k_y, k_z)$
   ensures $m_{\rm e}^2 >0$ 
   (see Equation~\eqref{eq_def_kappaie}). 
The ``improper'' part incorporates improper modes, which possess any frequency 
   satisfying $m_{\rm e}^2 <0$ and hence necessitate an integration over frequency.
Regardless, the ``lateral'' component in Figure~\ref{fig_vxAnim} is attributable
   to the improper contribution. 
      
We now consider the guided component by connecting Equation~\eqref{eq_vx_IVPformal}
   with the 
   resistive EVP results encapsulated in Figures~\ref{fig_vRainbow}
   and~\ref{fig_vgContour}. 
Before proceeding, however, the following remarks are necessary. 
Firstly, the group velocity $\vec{v}_{\rm gr}$ involves only the 
   real part ($\omega$) of the eigenfrequency $\Omega$. 
What a non-vanishing $\gamma=\Im\Omega$ means for an ideal computation
   is that the fast wave energy is transferred to localized \Alfvenic\
   motions where the \Alf\ resonance takes place, as has been extensively
   demonstrated for kink modes in cylindrical equilibria
   \citep[e.g.,][]{2008ApJ...687L.115T,2010ApJ...711..990P,2015ApJ...803...43S}.
The same also happens here in that \Alfvenic\ motions can be readily discerned
   as enhanced velocity shear $\partial v_y/\partial x$
   in some moving volumes in the slab boundary that accompany the strongest $v_x$ perturbations.   
Secondly, strictly speaking, our EVP analysis needs to account for
   the finite beta to ensure a more self-consistent application
   of the EVP computation to the IVP study.
On top of that, implied by such an application
   is that resonant absorption does not significantly impact the oscillation 
   frequency $\omega$, which may not hold in general
   \citep{2015ApJ...803...43S}.
Nonetheless, two reasons make us believe that this is not too serious an issue, 
   one being that $|\gamma/\omega|$ is consistently $<0.1$ for the wavevectors
   examined in Figure~\ref{fig_vgContour}, the other being related to some
   quantitative analysis in what follows.

\begin{figure}
\centering
\includegraphics[width=1.\columnwidth]{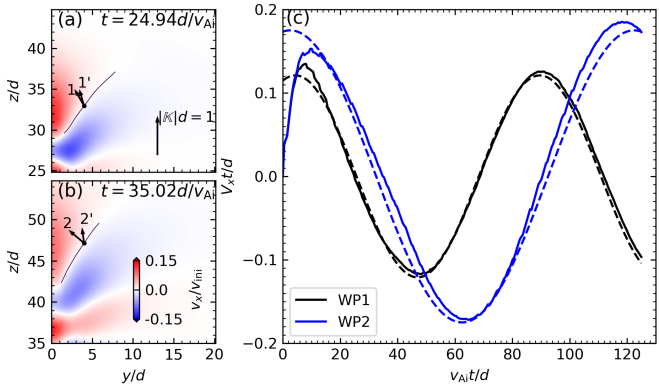}
\caption{
Left: Snapshots of $v_x$ 
   in the $x=0$ plane at two labeled instants. 
The solid curve in each panel is part of the outermost
   $v_x=0$ contour,
   the arrows representing the central wavevectors
   of the two wavepackets expected 
   for the location chosen on this iso-phase curve. 
The vertical arrow measures the magnitude of any wavevector.
Right: The time sequence of $v_x t$ sampled by an observer 
   moving with wavepacket $1$ (the black solid curve) 
   or wavepacket $2$ (blue).
Each sequences is fitted with a sinusoid 
   incorporating expectations from linear theory (the dashed curve).
See text for details.        
 }
\label{fig_MSP} 
\end{figure}

Figures~\ref{fig_MSP}a and \ref{fig_MSP}b present two zoomed-in $x=0$ cuts
   of $v_x$. 
A portion of the outermost $v_x=0$ contour is given. 
With this iso-phase curve we illustrate
   the relevance of Figure~\ref{fig_vRainbow}d to the large-time
   interference pattern.
Key is that the method of stationary phase (MSP) is increasingly applicable
   as time proceeds.
Suppose that the MSP applies to some $(y,z,t)$. 
Equation~\eqref{eq_vx_IVPformal} is then dominated by those wavepackets (WPs)
   with central wavevectors $\vec{K}_n=K_{n,y}\uvec{y}+K_{n,z}\uvec{z}$
   (\citetalias{1974Book..Whitham}, Equation~(11.41)),
\begin{equation}
\label{eq_MSP_vxLargeT}
v_x(0,y,z,t)\sim t^{-1}\sum\limits_{n}\left[\mathcal{G}_n(K_{n,y},K_{n,z})
      \Exp{\imath\left(\omega_n t-K_{n,y}y-K_{n,z}z\right)}
      \right],
\end{equation}
   where $\omega_n=\omega_n(\vec{K}_n)$ with $\vec{K}_n$ a solution to
\begin{equation}
\label{eq_MSP_K}
\vgy(\vec{K}_n)=y/t,
\quad
\vgz(\vec{K}_n)=z/t.
\end{equation}  
One complication, however, is that Equation~\eqref{eq_MSP_K}  
   possesses two solutions even if one assumes the relevance
   of only those modes in Figure~\ref{fig_vgContour}.
This is illustrated in Figure~\ref{fig_MSP}a where 
   the arrows represent the central wavevectors of the 
   two WPs that solve Equation~\eqref{eq_MSP_K} 
   for a point with $y=4d$ on the $v_x=0$ curve. 
Note that $K_y<0$ (see the symmetry property
   following Equation~\eqref{eq_omega_formal}).   
Let these WPs be labeled $1$ and $1'$, and suppose that 
   the unprimed WP dominates.
Two consequences follow.
First, $\vec{K}_1$ is locally normal to the iso-phase curve. 
Second, the time sequence seen by WP~1,
   $V_x(t)\coloneqq v_x(0,\vgy(\vec{K}_1)t,\vgz(\vec{K}_1)t,t)$,
   eventually becomes $\sim t^{-1}\sin[\varpi(\vec{K}_1) t+\phi(\vec{K}_1)]$ 
   with  
\begin{equation}
\varpi(\vec{K}_1)=\omega(\vec{K}_1)-K_{1,y}\vgy(\vec{K}_1)-K_{1,z}\vgz(\vec{K}_1)
\end{equation}
   being a Doppler-shifted frequency and $\phi$ some phase angle. 
This second property applies to any WP, and is hence
   useful for assessing whether the MSP applies
   or judging whether a WP dominates. 
The $V_x(t)t$ sequence seen by WP~1 is presented by  
   the black solid curve in Figure~\ref{fig_MSP}c. 
A sinusoid fitting with the expected $\varpi(\vec{K}_1)$ (the black dashed curve)   
   strongly suggests that the MSP applies to the trajectory of WP~1 when
   $t\gtrsim 7d/\vai$, despite the minor deviation of $\vec{K}_1$ from
   the local normal. 
The same practice is repeated for WP~2 in Figure~\ref{fig_MSP}b pertinent to
   a point with $y=4d$ on the solid curve.
Now the MSP applies for $t\gtrsim 10d/\vai$
   (see the blue curves in Figure~\ref{fig_MSP}c), 
   and $\vec{K}_2$ is nearly perfectly perpendicular to the iso-phase curve. 
The negative $y$-propagation of the delineated $v_x=0$ front
   may therefore be accounted for by the dispersion features in Figure~\ref{fig_vRainbow}d.

\section{Summary}
\label{sec_conc}
This study examined the three-dimensional (3D) propagation of kink wave trains
     impulsively excited by a localized perturbation
     to straight, field-aligned, symmetric, coronal slabs
     that are structured only in one transverse direction.
Two features stand out in our linear EVP analysis on oblique kink modes,
     namely the group and phase velocities may
     lie astride the equilibrium magnetic field ($\vec{B}_0$), 
     and the group trajectories extend only to a limited angle from $\vec{B}_0$.
These features were demonstrated numerically 
     for continuous and step structuring alike,
     and were understood with the approximate analytical expressions enabled
     by the latter.
Our 3D time-dependent simulation showed that these features are reflected
     in the intricate interference patterns at large times, the key being the ideas
     behind the method of stationary phase.
The guided wave trains in the out-of-plane cut through the slab axis
     are confined to a narrow sector about $\vec{B}_0$,
     with some wavefronts propagating toward $\vec{B}_0$. 

\section*{Acknowledgements}
We thank the referee (Dr. Roberto Soler) for constructive comments. 
This research was supported by the 
    National Natural Science Foundation of China
    (41974200, 41904150, and 11761141002).
We gratefully acknowledge ISSI-BJ for supporting the international team
    ``Magnetohydrodynamic wavetrains as a tool for probing the solar corona''.

\section*{Data Availability}
The data underlying this article are available in the article and in its references.



\bibliographystyle{mnras}
\bibliography{seis_generic} 








\bsp	
\label{lastpage}
\end{document}